\pdfoutput=1

\newif\iftwocol

\documentclass[final,5p,times,twocolumn,authoryear]{elsarticle}
\twocoltrue

\usepackage{lineno}
\usepackage{amssymb}
\usepackage{textcomp}
\usepackage{wasysym} 

\newcommand{\um}{\ensuremath{\,\mu{\rm m}}}

\newcommand{\methane}{\ensuremath{{{\rm CH}_4}}}
\newcommand{\be}{\begin{equation}} 
\newcommand{\ee}{\end{equation}}
\newcommand{\pmil}{\textperthousand\enspace}

\journal{Icarus}
\begin{document}
\begin{frontmatter}

\title{Search for methane isotope fractionation due to Rayleigh distillation on Titan}

\author[label1]{M\'at\'e~\'Ad\'amkovics}
\ead{mate@berkeley.edu}
\author[label2a,label2b]{Jonathan~L.~Mitchell}
\ead{jonmitch@g.ucla.edu}

\address[label1]{Astronomy Department, University of California, Berkeley, CA 94720, USA} 
\address[label2a]{Department of Earth, Planetary and Space Sciences, \\
University of California, Los Angeles, CA 90095, USA}
\address[label2b]{Department of Atmospheric and Oceanic Sciences, \\
University of California, Los Angeles, CA 90095, USA} 

\begin{abstract}

We search for meridional variation in the abundance of CH$_3$D relative to CH$_4$ on Titan
using near-IR spectra obtained with NIRSPAO at Keck, which have a photon-limited signal-to-noise
ratio of $\sim$50. Our observations can rule out a larger than 10\% variation in the column of
CH$_3$D below 50\,km. The preferential condensation of the heavy isotopologues will fractionate
methane by reducing CH$_3$D in the remaining vapor, and therefore these observations place limits on
the amount of condensation that occurs in the troposphere. While previous estimates of CH$_3$D
fractionation on Titan have estimated an upper limit of -6\textperthousand, assuming a solid
condensate, we consider more recent laboratory data for the equilibrium fractionation over liquid
methane, and use a Rayleigh distillation model to calculate fractionation in an ascending parcel of
air that is following a moist adiabat.  We find that deep, precipitating convection can enhance the
fractionation of the remaining methane vapor by -10 to -40\textperthousand, depending on the final
temperature of the rising parcel. By relating fractionation of our reference parcel model to
the pressure level where the moist adiabat achieves the required temperature, we argue that the
measured methane fractionation constrains the outflow level for a deep convective event.
Observations with a factor of at least 4 to 6 times larger signal-to-noise are required to detect
this amount of fractionation, depending on the altitude range over which the outflow from deep
convection occurs.

\end{abstract}


\end{frontmatter}


\section{Introduction}\label{s:Intro}

Titan's hydrological cycle is an interesting combination of atmospheric and surface processes, and
it is important for understanding the climate; for example, see the review by \citet{ML16}. Unlike
on Earth, where the water of the hydrological cycle is primarily in the condensed phase on the
surface, Titan's supply of methane is stored mostly in the atmosphere as vapor \citep{Lorenz2008}.
The column of methane vapor is equivalent to 5\,m of liquid at the surface \citep{Tokano2006a}.  
The large atmospheric reservoir contributes to the fact that the phase transitions of methane are a
significant part of the energy transport in the atmosphere \citep{Mitchell2009b}. The evaporation,
circulation, and condensation of atmospheric methane redistributes the energy that is input as short
wavelength solar radiation near to the equator, and moves it such that the outgoing long wavelength
radiation at the top of the atmosphere is emitted nearly isotropically \citep{Mitchell2012}.

Condensation and evaporation are critical for determining which regions of the surface are dry or
wet, with net evaporation drying the equatorial regions \citep{Mitchell2008, Lora2015a}. While
axisymmetric models of circulation predict the preponderance of ponding near the poles, they do not
explain why the northern polar surface has significantly more and larger lakes than in the south
\citep{Stofan2007,Hayes2008,Turtle2009}. One explanation for the asymmetric distribution of polar
lakes on Titan is Saturn's orbital eccentricity, which is thought to drive the seasonally-averaged
preferential evaporation from the south \citep{Aharonson2009}. Tropospheric eddies, which models
suggest are more vigorous in the south, are important for this process, as they pump the moisture
that is evaporating from the surface up to higher altitudes and towards lower latitudes, and supply
vapor to the upper atmosphere where the meridional overturning circulation takes place
\citep{Lora2015b}. Measuring the condensation of methane in the atmosphere, and how it 
changes with time, is one way to evaluate whether this explanation of the distribution of lakes is 
correct.

Indications of precipitation \citep{Perron2006,Tokano2006a,Adamkovics2007,Adamkovics2009,
Mitchell2011,Turtle2011a} and the observations of clouds \citep{Brown2002,Roe2002a,Roe2012}
demonstrate that condensation is occuring in the atmosphere, but measurements of methane vapor are
challenging. The Cassini-Huygens probe gas chromatograph mass spectrometer (GCMS) is the canonical
measurement of methane \citep{Niemann2010}, and is often used as the point of reference for
measurements at other locations and at other times. Cassini/CIRS spectra in the thermal IR are
sensitive to both the methane abundance and temperature in stratosphere, and the mole fraction
measured by the GCMS is used to retrieve thermal profiles
\citep[e.g.,][]{Flasar2005,Achterberg2011}. \citet{Lellouch2014} suggest that CIRS spectra can be
used to measure the stratospheric methane abundance, since the mid- and far-IR lines of methane have
different sensitivities to the thermal profile, and they find mole fractions near $\sim$85\,km
altitude of $\sim$1.0\% at low latitudes, contrary to the GCMS. At 70 N, where using the CIRS lines
to constrain the temperature is challenging, \citet{Lellouch2014} find a mole fraction of
$\sim$1.0\%, whereas \citet{Anderson2014} interpret the same observations with a temperature profile
that was independently determined with Cassini Radio Science Subsystem (RSS) and find a mole
fraction of $\sim$1.5\%. Near-IR measurements made with the Upward Looking Infrared Spectrometer
(ULIS) of the Descent Imager/Spectral Radiometer (DISR) are consistent with the GCMS, and it has
been suggested that the discrepancy with CIRS is due to uncertainties in methane line parameters
\citep{Bezard2014}. On the other hand, the near-IR measurements of methane abundance are sensitive
to properties of the atmospheric aerosol
\citep{Penteado2010b,A16}. A method of measuring condensation that does not necessarily rely on the
mole fraction would be a valuable complement to these techniques.

Condensation and evaporation impart an isotopic signature on a system because the rates of chemical
and physical processes are different among isotopologues, due to the differences in mass, bonding,
and zero point energies. In a closed system at equilibrium, the vapor pressures of different
isotopologues are a measure of the difference in these rates. For a physical system like a parcel of
air, where the condensate can be removed from the system, the fractionation can be larger still, as
the fractionated condensate leaves the system. Rayleigh fractionation is therefore a signature of
the magnitude and temperature at which evaporation and condensation occur, and measurements of
$^{18}$O and D in water are commonly used in interpreting properties of Earth's hydrological cycles
\citep{Dansgaard1964}.  Isotopic fractionation of condensible liquids is of considerable interest not only for
characterizing the terrestrial water cycle \citep[e.g., see review by][]{Xi2014}, but also for
constraining the Martian water cycle \citep[e.g.][]{Montmessin2005,Villanueva2015, Encrenaz2016}.
Similarly, the hydrological cycle on Titan can be informed by measurement of the isotopic
composition of methane.

Recent measurements of the methane humidity in the lower atmosphere were made using
spatially-resolved spectra in the 1.5\um\ (H-band) spectral region \citep{A16}. Here we present
further study of these spectra in the Section 2, focusing on a detailed uncertainty analysis. The
H-band is sensitive to two isotopologues of methane, offering the possibility of detecting variation
in the strength of CH$_3$D spectral features relative to those of CH$_4$. While systematic
instrumental noise and uncertainties in the gas-phase opacity are important, searching for variation
in CH$_3$D relative to CH$_4$, and in spectra at one location relative to another, means that we can
take into account the systematic effects. In Section 3 we describe the 
S/N
required to
measure a given magnitude of fractionation, and in Section 4 we discuss these results interpreted
with a Rayleigh fractionation model, where we estimate the magnitude of fractionation in a parcel of
air with condensation.

\iftwocol
	\begin{figure*} \includegraphics[width=7in]{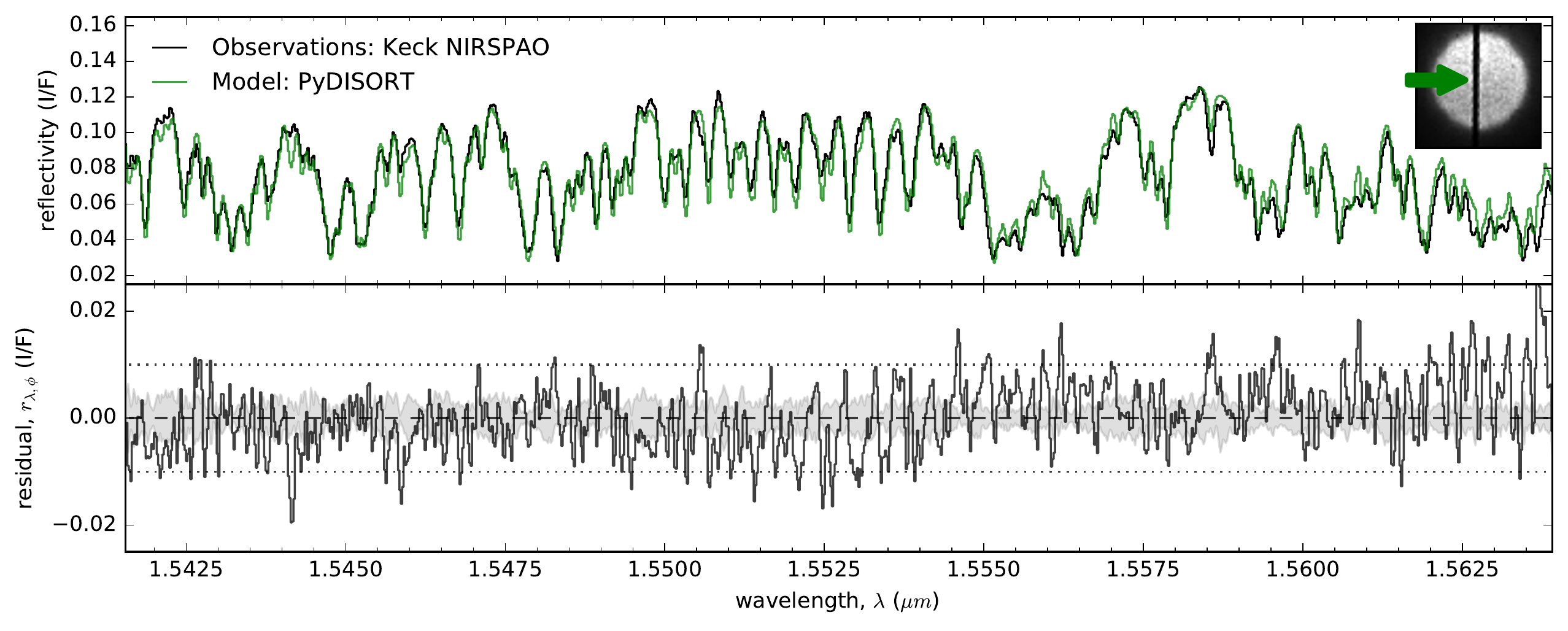}  
\else
	\begin{figure} \includegraphics[width=5.5in]{example_spectrum.pdf}  
\fi
	\begin{center}
	\caption{\label{f:obs} The spectrum from one NIRSPAO pixel (top panel; black), at the location
	specified by the green arrow in the slit-viewing camera image (inset), is compared with the
	radiative transfer model spectrum (green). The residuals are plotted in the bottom panel, with
	the ordinate scale magnified by a factor of three relative to the top panel. The shaded grey
	region is the standard deviation of the observed flux from the neighboring five pixels, which
	gives an estimate of the pixel-to-pixel noise in the observations.}
	\end{center} 
\iftwocol
	\end{figure*}
\else	
	\end{figure}
\fi

\section{Methods}\label{s:methods}

\subsection{Observations}

The Near-InfraRed SPECctrometer \citep{McLean1998} with adaptive optics (NIRSPAO) was used at W.~M.
Keck Observatory on 17~July~2014~UT to observe Titan with a spectral resolving power of
$R\approx25,000$ and a spatial sampling  of 0.018"/pixel along the slit. A single North-to-South
position along the central meridian was integrated for 45min. We analyze spectra from one echelle
order centered near 1.55\um. Additional details of these observations, including the data reduction
and calibration with supporting datasets, are described in \citet{A16}.

\iftwocol
	\begin{figure*} \includegraphics[width=7in]{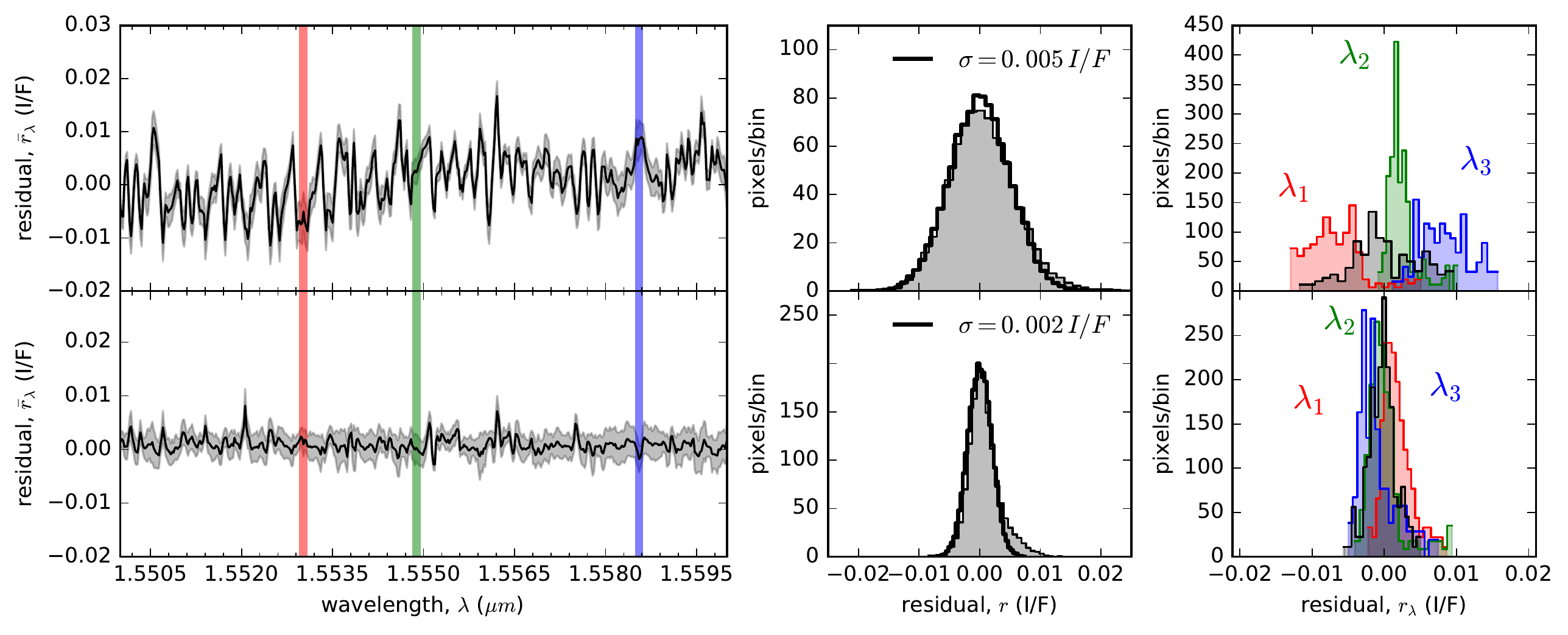} 
\else
	\begin{figure} \includegraphics[width=5.5in]{residuals.pdf}
\fi
	\begin{center}
	\caption{\label{f:residuals}
	Residuals for the standard model (top row of panels) are chacterized by variations in
	$\bar{r}_{\lambda}$ (black line) that are larger than $\sigma_{\lambda}$ (grey shaded region).
	The center panel is a histogram of the residuals $r$ for the entire sample (grey) shown with a
	histogram for a random normal distribution of the same sample size with $\sigma=0.005$.
	Residuals from wavelengths corresponding to the vertical lines of matching color are plotted in
	the right panel, which can be discrepant from a histogram of a random noise (black). The bottom
	row of panels is for residuals from models that include an empirical correction for the
	systematic error (see text for details). After correction, the structure in the spectrum of
	$\bar{r}_{\lambda}$ is removed (left panel), $r$ is consistent with random noise with
	$\sigma=0.002$  (center panel), and $r_{\lambda}$ is distributed around zero for individual
	wavelengths (right panel).}
	\end{center} 
\iftwocol
	\end{figure*}
\else	
	\end{figure}
\fi

\subsection{Radiative Transfer Model}

Synthetic spectra are generated by defining 20 atmospheric layers, with properties that are
determined primarily by measurements made with instruments on the {\it Huygens} probe. The layers
have boundaries (levels) that are evenly spaced in pressure, with 10 levels above and 10 levels
below 300 mbar \citep[see Table~2 in][]{A16}. The top of the atmosphere is set at zero optical
depth. We use measurements of the temperature, pressure, methane abundance, and aerosol structure to
determine the gas and scattering opacity in each layer. The methane below 35\,km and aerosol
above 65\,km are assumed to vary with latitude, as detailed in \citet{A16}. Methane is
increasing linearly from 40N toward the south, reaching a 30\% enhancement at 40S, while the aerosol
is increasing in opacity toward the north. The {\it Huygens} temperature profile is used at all
latitudes. CH$_4$ and CH$_3$D line opacities are from the HITRAN 2012 database \citep{Rothman2013}.
The discrete-ordinate-method radiative transfer\citep[DISORT;][]{Stamnes1988} is implemented in
Python (PyDISORT) and used to solve the radiative transfer through the model atmosphere and simulate
the observed flux.

Each ($x,y$) pixel on the detector maps to a wavelength and latitude ($\lambda_x, \phi_y$) for the
observed flux, $I_{\rm obs}(\lambda_x,\phi_y)$. We use the radiative transfer model to calculate the
flux $I_{\rm calc}(\lambda_x,\phi_y)$ corresponding to this wavelength and location, taking into
account the viewing geometry and spatial variation in surface albedo and hazes \citep{A16}. An
example spectrum from one NIRSPAO pixel along the slit (out of a total of 44 that cover the disk of
Titan), is compared with the radiative transfer model calculation in Figure~\ref{f:obs}.

\subsection{Uncertainty Analysis}

By inspecting a 300s exposure, we find the background count rate at echelle Order 49 to be 0.21
DN/s, while the count rate on Titan ranges from 0.08 to 0.63 DN/s, in the dark and bright spectral
regions, respectively. Using a NIRSPAO dark current of 0.7e$^-$/s/pix, gain of 5.7e$^-$/DN, read
noise of 23e$^-$, and total 9$\times$300s exposure time, the counting-limited signal-to-noise ratio
S/N per pixel ranges from 19 in the dark regions to 87 where Titan is bright. We'll use the mean
value to characterize the entire spectrum as S/N$\approx$50. In units of reflectivity, which cover a
range of roughly 0.03 to 0.12$\,I/F$, this level of noise corresponds to $\sim$0.0015$\,I/F$ per
pixel and sets the limit for the expected residuals when comparing these observations to our models.

The residual for each pixel is defined with the following notation,
\be
r_{\lambda,\phi} = I_{\rm obs}(\lambda_x,\phi_y) - I_{\rm calc}(\lambda_x,\phi_y),
\ee
and we consider the standard deviation $\sigma$ for the entire sample of residuals $r$, as well as
sub-samples at a specific wavelength $r_{\lambda}$ or latitude $r_{\phi}$, which have mean residuals
and standard deviations of $\bar{r}_{\lambda}$, $\bar{r}_{\phi}$ and $\sigma_{\lambda}$,
$\sigma_{\phi}$, respectively.

The residuals that are shown in Figure~\ref{f:obs} are often in excess of the per pixel noise of the
observations determined above, as well as the pixel-to-pixel noise of the observations determined by
comparing nearby pixels. This indicates that systematic errors dominate and is supported by the
similarity of the residuals at other spatial locations (see, e.g,. Figure~5 in \citet{A16}). The
systematic error is illustrated with a spectrum of the mean residual for all spatial pixels,
$\bar{r}_{\lambda}$, in Figure~\ref{f:residuals}. For comparison, we generate a random normal sample
of Gaussian noise in a two-dimensional array that is the same size as the observations and compare
with the residuals. The entire sample $r$ is characterized by a normal distribution with
$\sigma=0.005\,I/F$, in excess of the estimated per pixel noise of the observations. The
distribution of residuals at individual wavelengths are different, and inconsistent with random
noise. At wavelengths where the systematic error is greatest, $|\bar{r}_{\lambda}| >
\sigma_{\lambda}$, indicating significant discrepancy.

\iftwocol
	\begin{figure*} \includegraphics[width=7in]{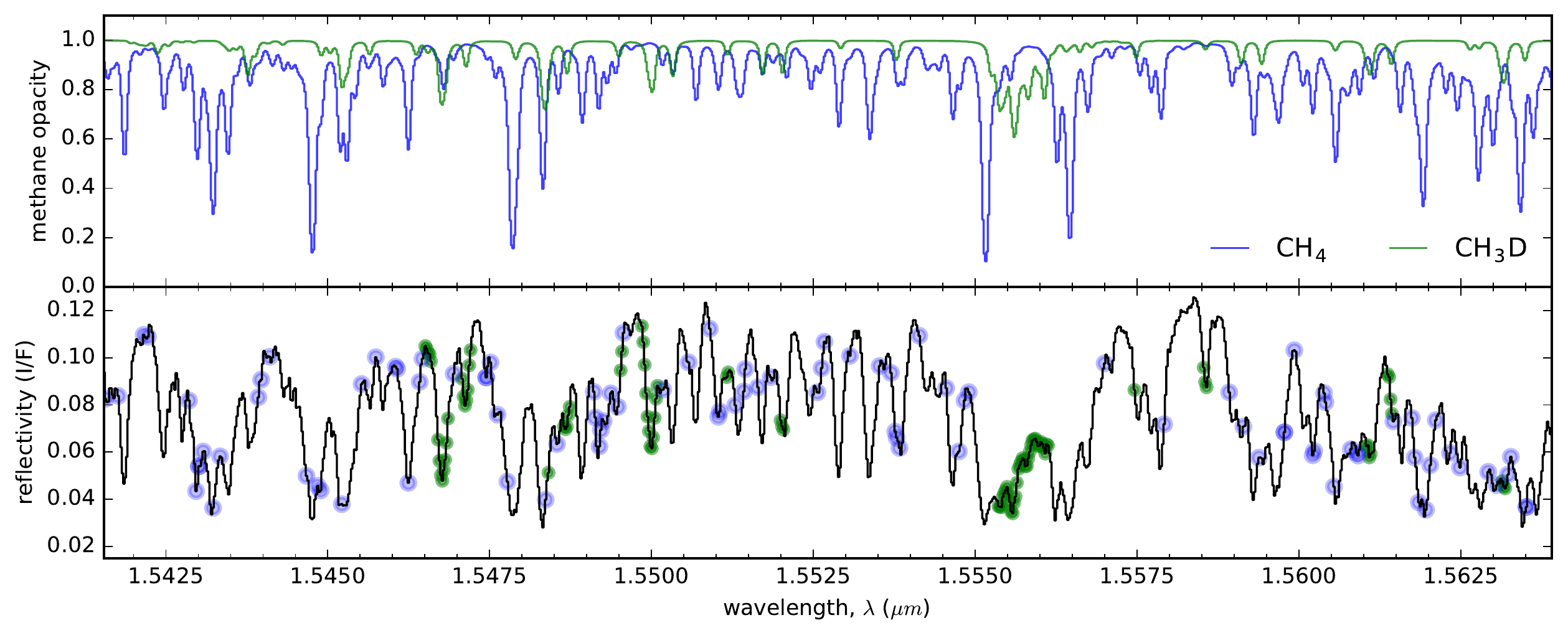} 
\else
	\begin{figure} \includegraphics[width=5.5in]{CH3D_windows.pdf}
\fi
	\begin{center}
	\caption{\label{f:CH$_3$D_windows}Transmission spectra of CH$_4$ (blue) and CH$_3$D (green) in
	 the top panel at a temperature $T=94$K, pressure $P=1$bar, and unit path length. The wavelength
	 regions where CH$_3$D dominates the methane gas opacity are highlighted with green circles in
	 the observed spectrum in the bottom panel. A random selection of the same number of datapoints
	 that are sensitive to CH$_4$ are shown in blue. These two sets of wavelengths are used to
	 search for spatial variation in CH$_3$D relative to CH$_4$. }
	\end{center} 
\iftwocol
	\end{figure*}
\else	
	\end{figure}
\fi

The systematic spectral error could arise from some combination of the following: either inadequate
removal of contamination (e.g., from scattered light) or other systematic instrumental noise during
the data reduction, or uncertainties in the gas phase opacities considered in the radiative transfer
model. The latter could involve either ignoring gas phase species that contribute to the spectrum, 
or limitations in the absorption coefficients. Since the features in the spectrum of
$\bar{r}_{\lambda}$ have widths that are comparable to the methane spectrum, the systematic error
may have a contribution from limitations in the gas opacity.

To compensate for the systematic error, we implement an empirical correction to the gas opacity.
Since methane dominates the gas opacity, we make an ad hoc correction to the methane opacities.
The corrected methane opacity spectrum $\tau_{\rm corr}(\lambda)$ is calculated using the standard
methane opacity $\tau(\lambda)$ and residuals $\bar{r}_{\lambda}$,
\be
	\tau_{\rm corr}(\lambda) = \tau(\lambda) \exp\left( c_0 \bar{r}_{\lambda} \right), 
\ee
where the coefficient $c_0$ is set to minimize the residuals in the models using the corrected
opacities (in this case $c_0=60$). The calculation of $\tau(\lambda)$ is described in \citet{A16}.
The same correction factor, $\exp\left( c_0 \bar{r}_{\lambda} \right)$, is applied to all layers in
the model and to all latitudes. Residuals from a model that used these corrected opacities are shown
in the bottom row of Figure~\ref{f:residuals}, with $\sigma=0.002\,I/F$, which is close to the
photon-limited per pixel S/N of 0.0015$\,I/F$. A tail in the distribution of $r$ to large positive
values (and spikes in the spectrum of $\bar{r}_{\lambda}$) still exist. This ad hoc correction to
the model consolidates all of the systematic uncertainty into the methane opacity, and because there
are correction factors of up to a factor of 2, it is unlikely that there are inaccuracies of this
magnitude in the laboratory data. If future observations show a similar spectrum of systematic
noise, especially if the instrument grating and cross-disperser angles are different, resulting in
the spectra falling on different pixels, then it may suggest that the methane opacities could indeed
be the culprit; however, if the systematic noise is different, it will point to an instrumental
artifact.

\section{Results}\label{s:Results}

\begin{figure}
\begin{center}
\iftwocol		
	\includegraphics[width=3.5in]{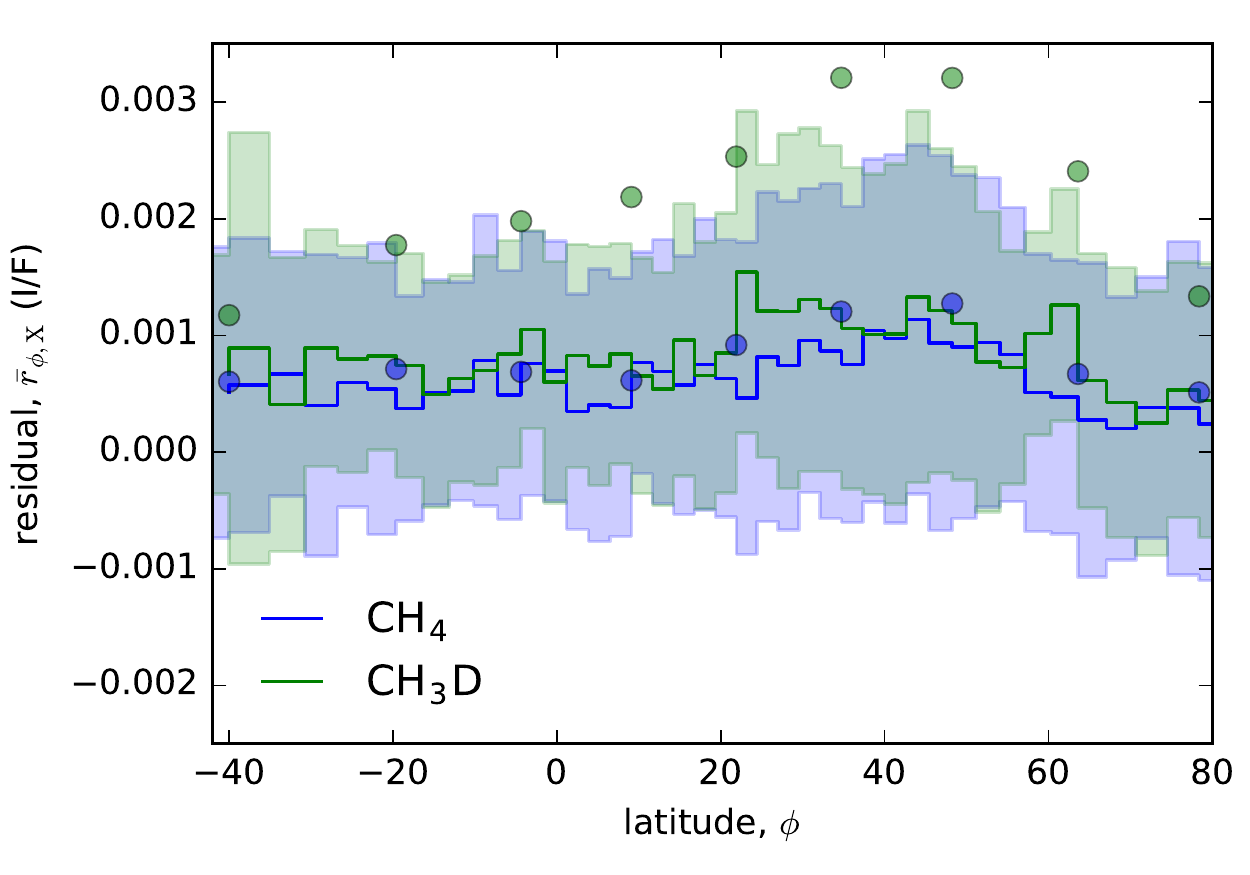}
\else
	\includegraphics[width=3.5in]{CH3D_search.pdf}  
\fi
	\caption{\label{f:CH$_3$D_search}Meridional trends in the median residual at wavelengths
	that are preferentially sensitive to CH$_3$D (green) and \methane\ (blue). The shaded regions
	are the uncertainty $\sigma_{\phi, {\rm X}}$ for the sample at each latitude. Residuals in a
	model where CH$_3$D is decreased by 10\% are shown as circles, indicating that a reduction of
	this magnitude is detectable when considering these wavelengths. }
\end{center} 
\end{figure}

There are several wavelength regions (or channels) near 1.5\um\ that are sensitive to CH$_3$D.
Transmission spectra for both isotopologues are shown in the top panel of
Figure~\ref{f:CH$_3$D_windows} and the wavelength regions  where the CH$_3$D opacity exceeds that of
CH$_4$ by at least 15\% are highlighted (103 channels altogether). Collectively, this sample of
wavelengths covers a range of $I/F$, and are therefore sensitive to CH$_3$D at various
altitudes in the atmosphere. We randomly sample 103 channels from the remaining $\sim$900 that are
sensitive to CH$_4$ for a baseline comparison. The meridional trend in the mean of the residuals for
these two sets of channels $\bar{r}_{\phi, {\rm X}}$, where X specifies CH$_3$D or \methane, are
plotted in Figure~\ref{f:CH$_3$D_search}, showing that no variation in CH$_3$D relative to CH$_4$ is
detected.

To determine what our sensitivity is with a per pixel noise of 0.002$\,I/F$, we first consider a
10\% decrease in CH$_3$D relative to CH$_4$ below 50\,km altitude. The datapoints in
Figure~\ref{f:CH$_3$D_search} illustrate that decreasing the abundance of CH$_3$D by 10\%, while
keeping CH$_4$ fixed, causes a model residual outside the uncertainty over a significant fraction of
the disk. There is increased sensitivity near and sub-observer point where the 10\% decrease causes
the larger residuals that occur near the limb. There are also large residuals toward Northern
latitudes where the surface albedo is higher and the returned flux is greater. As a comparison, the
set of baseline wavelengths, which are more sensitive to CH$_4$, show little variation in the
residuals when only the abundance of CH$_3$D is changed. The systematic error is a factor of
2.5 larger for the entire sample (see Figure~\ref{f:residuals}). If the empirical correction is not
included, this corresponds to a sensitivity to a 25\% decrease in CH$_3$D.

To generalize the sensitivity to changes in CH$_3$D  abundance for a particular S/N, we compare
models with a range of CH$_3$D abundances to our reference model. We use the standard
notation for isotopic variation,
\be
\delta =  (a-a_{\rm std})/a_{\rm std} \times 10^3\,\permil
\ee
where the absolute content $a$ of an isotope is compared to a standard reference value $a_{\rm
std}$. For the two isotopologues of methane we consider specifically
\be
\delta\, {\rm CH_3D} = {\rm \frac {\left( \frac{CH_3D}{CH_4} \right) - 
                              \left( \frac{CH_3D}{CH_4} \right)_{std}}
                             {\left( \frac{CH_3D}{CH_4} \right)_{std}}
                       }  \times 10^3\,\permil,
\ee
where were use the terrestrial D/H=1.56$\times 10^{-4}$ for our standard ${\rm (CH_3D/CH_4)_{std}}$,
which is consistent with the isotope ratio observed on Titan \citep{Nixon2012}.

\begin{figure}
\begin{center}
\iftwocol		
	\includegraphics[width=3.5in]{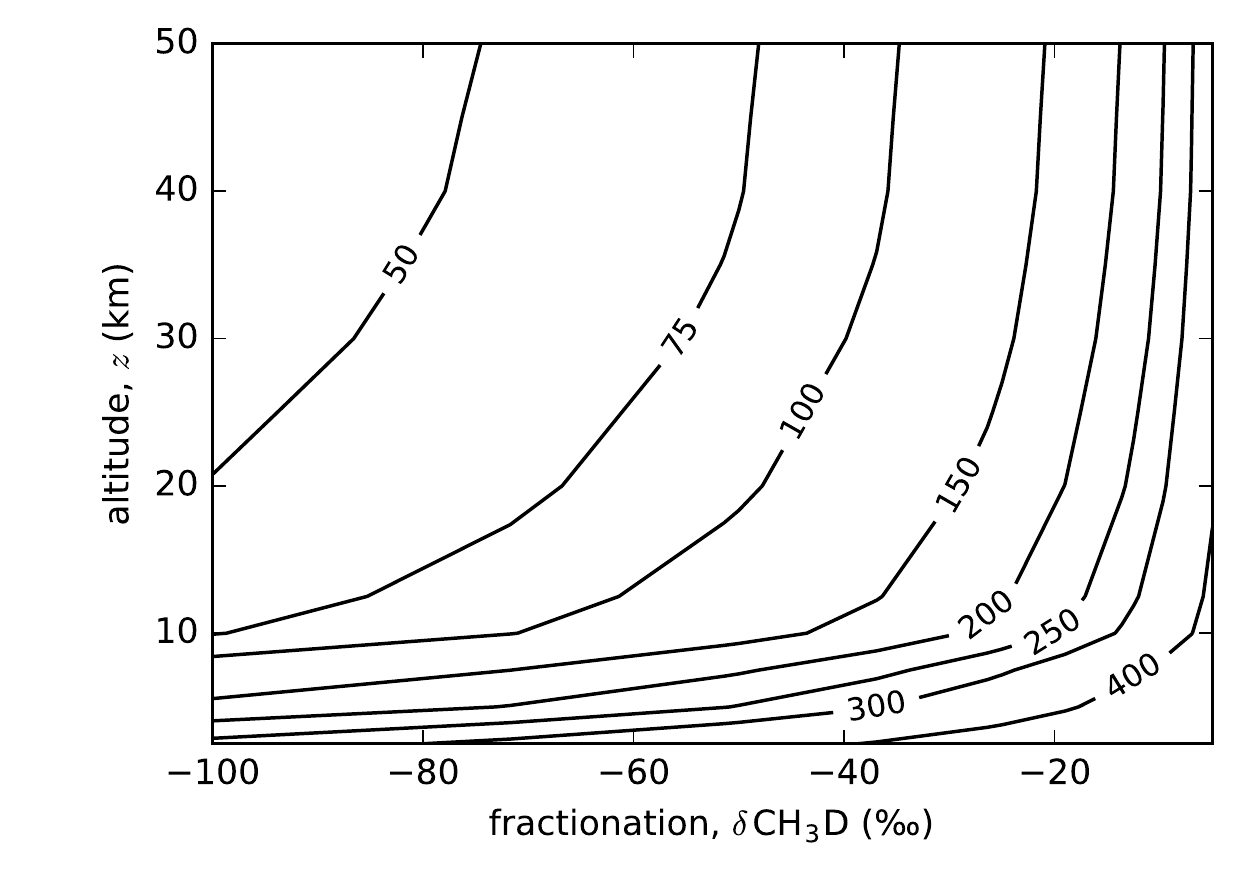}
\else
	\includegraphics[width=3.5in]{SN_contours.pdf}  
\fi
	\caption{\label{f:SN_contours}The per pixel S/N that is required to observe a 
	$\delta {\rm CH_3D}$ of fractionation below a given altitude $z$ in the atmosphere.
	}
\end{center} 
\end{figure}

We calculate the mean residual $\bar{r}_{\delta}$, between the standard calculation and one with
$\delta\, {\rm CH_3D}$ depletion, over the sample of wavelengths that are shown to be sensitive to
CH$_3$D in Figures~\ref{f:CH$_3$D_windows} and \ref{f:CH$_3$D_search}. If the noise in the
observations is ${\rm N} < \bar{r}_{\delta}$, the variation of $\delta\, {\rm CH_3D}$ is
detectable.  By considering a characteristic signal of magnitude ${\rm S}=0.1\,I/F$, we can then
determine the limiting S/N required to measure a particular depletion. A contour plot of the
limiting S/N required for these two parameters is shown in Figure~\ref{f:SN_contours}.  Depletion
from below lower altitudes means that a smaller column of CH$_3$D is removed, so that the resulting
change in the spectrum is smaller, requiring a higher S/N to detect. Similarly, smaller depletions
at a given altitude require higher S/N.

\section{Discussion}\label{s:Discussion}

If condensation occurs under Rayleigh conditions, where the condensate is immediately 
removed from the vapor after formation, and we assume that process is isothermal 
at temperature $T$, then the isotopic fractionation for the vapor and condensate are
\be
\label{e:dv}
\delta_v = \frac{1}{\alpha}F_v^{\,\alpha-1} - 1
\ee
and
\be
\label{e:dc}
\delta_c = F_v^{\,\alpha-1} - 1,
\ee
respectively, where $\alpha$ is the fractionation factor (or fractionation coefficient)
and $F_v$ is the remaining fraction of vapor \citep{Dansgaard1964}. 
The fractionation factor is the ratio of vapor pressures for the two isotopologues. 

\begin{figure}
\begin{center}
\iftwocol		
	\includegraphics[width=3.5in]{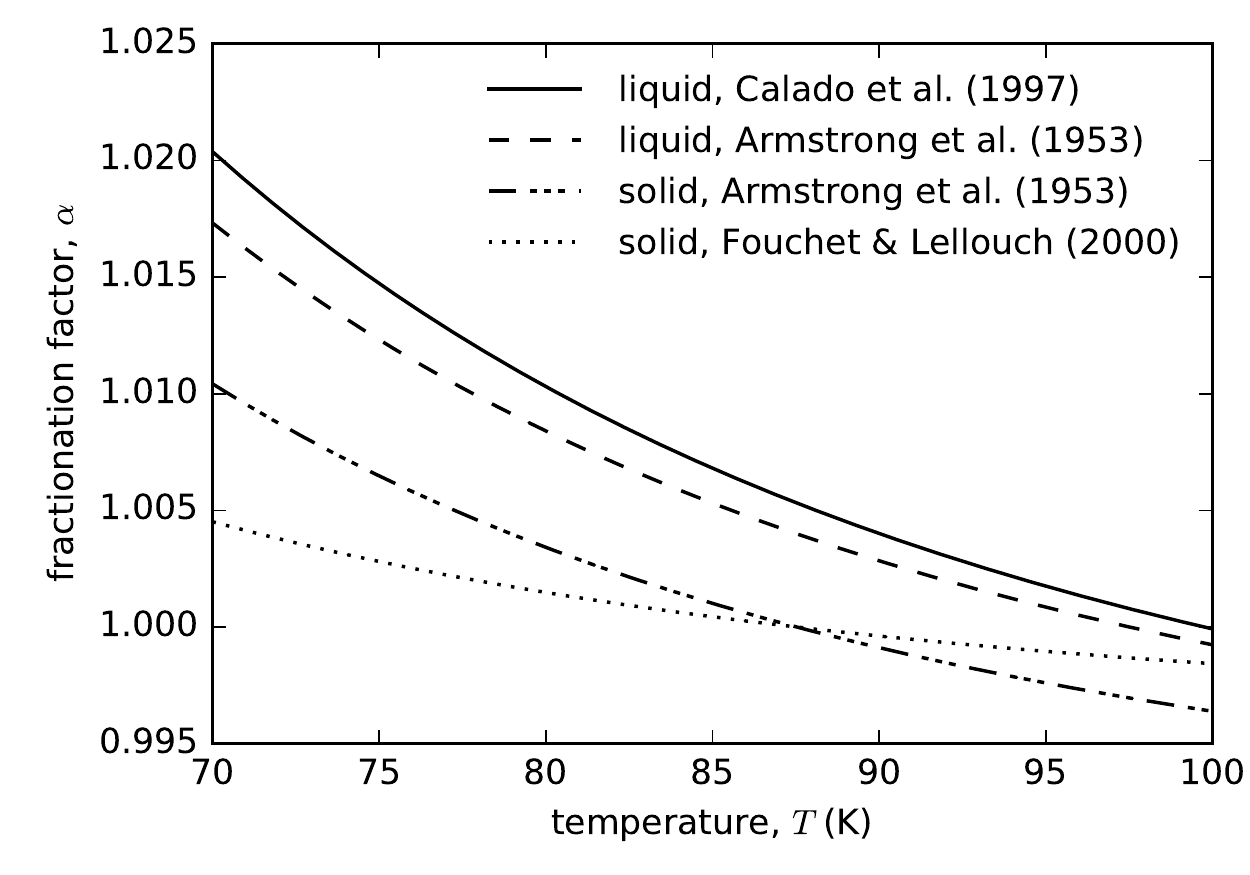}
\else
	\includegraphics[width=3.5in]{alpha.pdf}  
\fi
	\caption{\label{f:alpha}The fractionation factor, $\alpha$, for vapor in 
	equilibrium with solid and liquid methane from the two-parameter fits 
	to the laboratory data of \citet{Armstrong1953} and \citet{Calado1997}, 
	along with a curve for the expression used by \citet{Fouchet2000}.
	}
\end{center} 
\end{figure}

Laboratory measurements of the vapor pressures of methane isotopologues in equilibrium with solid
and liquid methane at different temperatures are reported with various two-parameters fits
\citep{Armstrong1953,Calado1997}, and are shown in Figure~\ref{f:alpha}. Also included is
the expression used by \citet{Fouchet2000}, where they assumed a maximum $\alpha=1.004$ at 71\,K, 
which they used to calculate a maximum -6\pmil fractionation on Titan for solid methane condensate.

Here we use the more recent laboratory measurements from \citet{Calado1997} for the 
liquid methane fractionation factor for CH$_3$D/CH$_4$,
\be
\alpha = \exp\left(\frac{A}{T^2}-\frac{B}{T} \right),
\ee
with $A=331.29\,{\rm K^2}$ and $B=3.3208$\,K. At 85\,K, $\alpha=1.0068$, and we can use
Equations~\ref{e:dv} and \ref{e:dc} to calculate the CH$_3$D depletion in the vapor that is expected
during methane condensation in a parcel where precipitation removes the liquid condensate. Curves
for the expected fractionation of the condensate and vapor during condensation are shown in
Figure~\ref{f:fractionation}. Since we make the simplifying assumption that condensation is
isothermal, we also calculate $\delta\, {\rm CH_3D}$ at 80\,K and 90\,K (left panel of
Figure~\ref{f:outflow}). These temperatures apply to the lower troposphere, up to an altitude of
$\sim$15\,km. Coincidentally, a fractionation factor of $\alpha\approx1.010$ for vapor over liquid
methane at 80\,K is roughly the same as for vapor over solid methane at 70\,K, which would be
applicable at higher altitudes, in the 30-50\,km range. Predictions for fractionation based on an
isothermal Rayleigh distillation model fall in the -5 to -40\pmil range, depending on temperature
and the fraction of vapor that condenses. While these estimates bracket a range of plausible values,
we next consider a more realistic scenario.

\iftwocol
	\begin{figure} \includegraphics[width=3.5in]{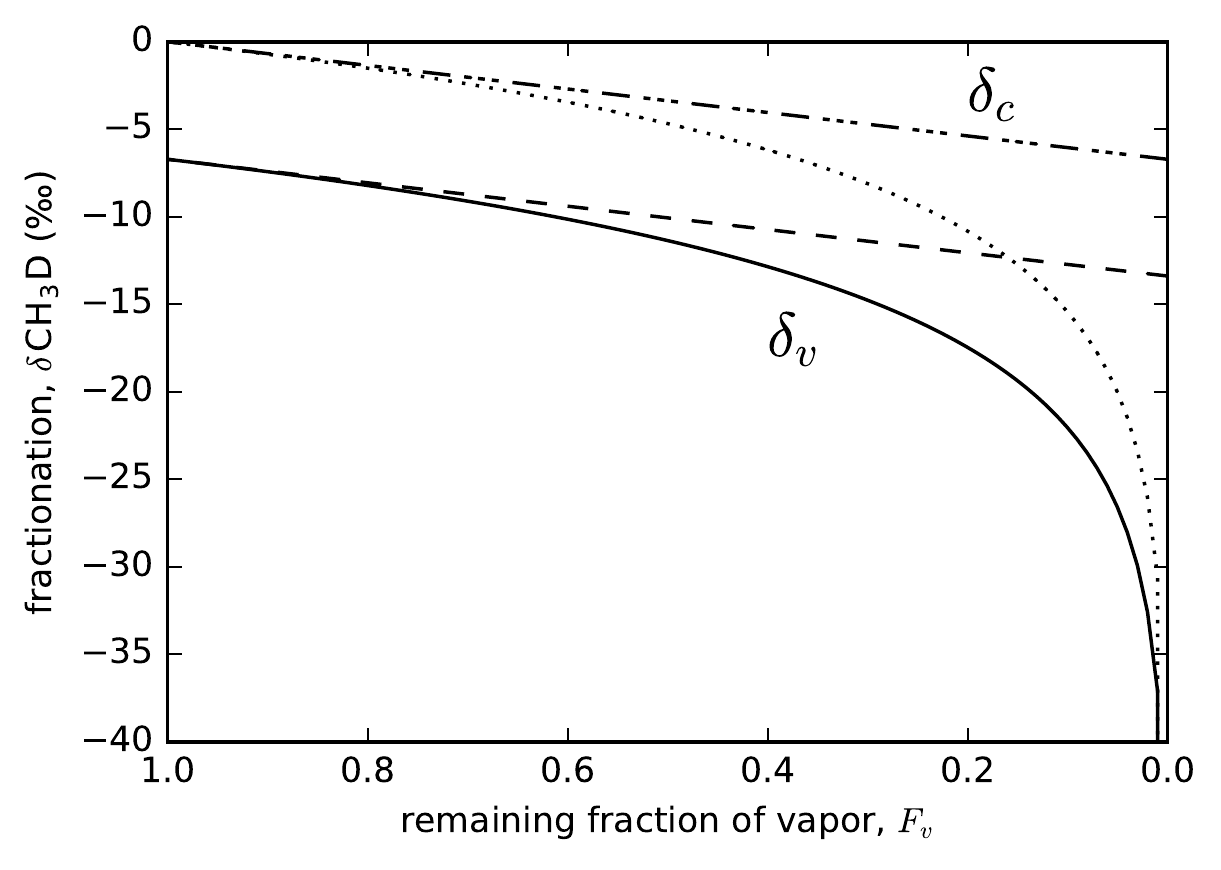} 
\else
	\begin{figure} \includegraphics[width=3.5in]{fractionation.pdf}
\fi
	\begin{center}
	\caption{\label{f:fractionation}The fractionation of methane in the remaining  
	vapor $\delta_v$ and in the condensate $\delta_c$, during condensation using an equilibrium
	model (dashed lines) and isothermal  Rayleigh distillition at 85\,K (solid and dotted curves),
	where the condensate is removed from the system.}
	\end{center} 
\iftwocol
	\end{figure}
\else	
	\end{figure}
\fi

Our working hypothesis is that methane vapor is supplied from near-surface environments to the upper
troposphere by localized, deep (precipitating) convection, such as the storms observed at the south
pole following the southern-hemisphere summer solstice \citep[e.g.,][]{Schaller2006a,Schaller2006b}
and near the equator at equinox \citep{Turtle2011c}.  We imagine a surface-level parcel of air that
is lifted beyond the condensation level following the moist adiabat.  The parcel therefore maintains
a saturated value of humidity, losing the excess to condensation of cloud droplets/ice and
precipitation.  In a vigorous updraft, the parcel reaches a stable, equilibrium level and is forced
to outflow laterally into the surrounding environment.  Because Titan has such weak lateral
temperature gradients, the parcel's humidity upon outflow is conserved.  Therefore the parcel
retains the isotope signal it inherited at the point of outflow from the convective updraft. Because
Huygens observed nearly saturated methane vapor above the boundary layer, it seems plausible that
many such parcels have mixed to essentially saturate the entire troposphere above the boundary
layer ($\sim$5-40km altitude) with methane.

\iftwocol
	\begin{figure*} \includegraphics[width=7in]{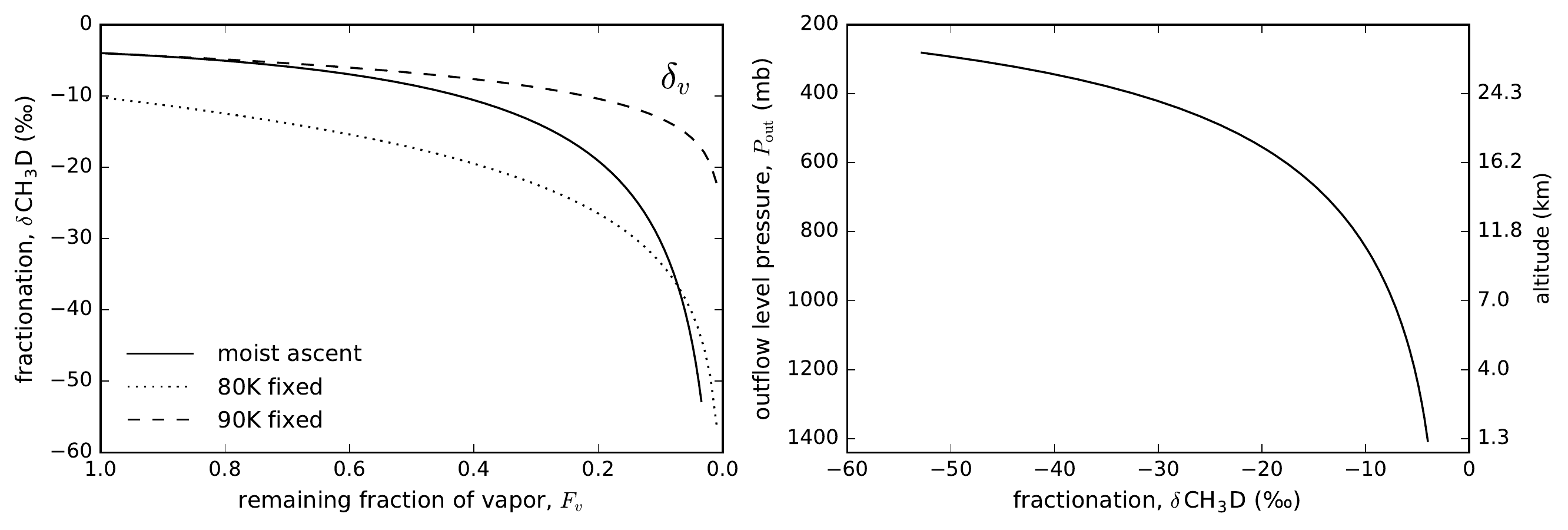} 
\else
	\begin{figure} \includegraphics[width=5.5in]{outflow_level.pdf}
\fi
	\begin{center}
	\caption{\label{f:outflow} The methane vapor fractionation for a parcel of air ascending along
	 the moist adiabat from the LCL (solid curve) is compared with isothermal fractionation at 80\,K
	 (dotted curve) and 90\,K (dashed) in the left panel. Since the temperature and remaining
	 fraction of vapor are known for a parcel ascending along the moist adiabat, the fractionation
	 in the vapor can be used to determine the outflow pressure level for a precipitating deep
	 convective event (right panel).}
	\end{center} 
\iftwocol
	\end{figure*}
\else	
	\end{figure}
\fi

With an estimate of the temperature and dew point temperature (or any other measure of methane vapor
concentration) of a near-surface parcel, our reference model of how methane vapor is supplied
to Titan's free atmosphere (all levels above the boundary layer) provides all the necessary
ingredients to relate changes in vapor fractionation with changes in outflow levels.  Here we
provide a sample calculation to illustrate the magnitude of fractionation changes that could be
present in Titan's methane vapor.  We start with the idealized Clausius-Clapeyron equation
\begin{equation} \label{eq:CC}
e_s(T) = e_{s_0} \exp \left[{H_{lv}\over R_v} \left({1\over T_0}-{1\over T}\right)\right]
\end{equation}
with triple-point pressure, $e_{s_0}$, and temperature, $T_0$, latent heat of vaporization, $H_{lv}$
and methane gas constant, $R_v$.  Alternative vapor pressure curves could be implemented, however
the magnitude of the uncertainty due to the choice of an idealized expression is negligible relative
to other uncertainties, such as the isotopic composition of the methane source region.  The lifting
condensation level (LCL) can be found by following a dry adiabat up from the surface parcel until
the dew point temperature is reached. Above this level, we assume the parcel maintains saturation by
condensing/freezing the excess and immediately removing the liquid/solid portion from the parcel.
The parcel fractionation at any level is the sum of all the incremental condensation that occurs as
the parcel is lifted to that level.  In other words, we must integrate,
\begin{equation}\label{eq:ddeltdT}
{d\delta_v\over dT} = {F_v^{\alpha-2}\over\alpha^2}\left[F_v{d\alpha\over dT}(\alpha\ln F_v -1)+ 
(\alpha -1)\,\alpha{dF_v\over dT}\right] ,
\end{equation}
over the temperature range experienced by the parcel during ascent along the moist adiabat.  
Equation~\ref{eq:ddeltdT} is found by differentiating the equation for $\delta_v(T)$, 
Equation~\ref{e:dv}, with respect to temperature.

The left panel of Figure~\ref{f:outflow} displays the results of this integration for a parcel with
a 90\,K dew point temperature starting at an LCL of $\sim$1400 mbar up to a level where the moist
adiabat reaches 70\,K, and as a function of the fraction of remaining vapor.  For comparison, the
fractionation curves for isothermal parcels at 80\,K and 90\,K are also shown. Because the parcel's
temperature is constrained to be on the moist adiabat, and because we know the fraction of remaining
vapor at saturation given the temperature, we can use the moist adiabat, $T_{ma}(p)$, to map the
value of $F_v$ onto the outflow level pressure, $P_{\rm out}$, of the parcel; this is shown in the
right panel of Figure~\ref{f:outflow}. From this figure we see that a deep, precipitating
convective event with an outflow level of 400 mbar will have fractionated the source parcel's
methane by $\sim$50 per mil. Since we do not know the parcel's initial isotopic composition, we
therefore do not necessarily know the correct CH$_3$D/CH$_4$ standard against which to measure the
fractionation, However, if large differences in methane vapor fractionation are observed in latitude
or altitude, we may infer that outflow from deep convection occurs at varying levels.  This is, in
fact, expected to be the case; polar convection is likely to reach higher altitudes (lower
pressures) than low-latitude convection, presumably because surface parcels have higher humidities
at high latitudes
\citep{Griffith2008}.

\citet{Nixon2012} describe measurements of isotopic composition made with the CIRS instrument on
the Cassini spacecraft, which are sensitive to stratospheric altitudes ($\sim$80 -- 280\,km),
finding that the D/H ratios in methane are consistent with the terrestrial value. They also
summarize previous measurements from the literature, including ground-based measurements in the
near-IR that are sensitive to a methane column that extends down to the surface
\citep{Penteado2005,deBergh2012}. It may be interesting to note that while the ground-based
measurements agree within uncertainties with the CIRS observations, they correspond to slightly
smaller D/H ratios than those measured in the stratosphere. Determining whether or not this is due
to actual altitude or temporal variation in the isotopic composition, rather than systematic
differences among the measurements, will require higher signal-to-noise observations with smaller
measurement uncertainties.

Recent {\it ab initio} calculations have been used to make additional rovibrational assignments of 
CH$_4$ lines and sub-bands \citep{Rey2016}, which can be used to (re)evaluate the
uncertainties in the methane opacities used in our model. While implementing and critically 
evaluating these recent assignments is beyond the scope of this work, future observations at 
higher S/N, together with future assignments of currently unassigned CH$_3$D lines, will provide
a means to reduce the systematic uncertainties in the analysis presented here.

In summary, we have revisited high resolution NIRSPAO observations of Titan searching for a spatial
variation in the isotopic ratio of D/H in methane. We can exclude variation in CH$_3$D that is
greater than 10\% with the current observations. We describe isothermal models of Rayleigh
distillation, which predict that fractionation of -10 to -40\permil\, should occur for a condensing
parcel of air, and we show that fractionation of this magnitude can be used to determine the outflow
level for a deep convective event. We predict that NIRSPAO observations with a factor of 4 to 6
higher S/N are required to measure depletions of this magnitude. Such observations will be
facilitated by the NIRSPEC instrument upgrade \citep{Martin2014}, and will be feasible with the next
generation of ground-based telescopes such as the Thirty Meter Telescope \citep{Skidmore2015} and
the European Extremely Large Telescope \citep{Tamai2014}.

\section*{Acknowledgements}
This work was supported by NASA grants NNX14AG82G and NNX12AM81G, and the manuscript was
improved with the constructive comments of two anonymous reviewers.

\section*{References}

\bibliographystyle{elsarticle-harv} 
\bibliography{refs}

\end{document}